\documentstyle[12pt,epsf]{article}
\begin{document}

\thispagestyle{empty}
\begin{flushright}
SLAC-PUB-7381\\
hep-ph/9612307\\
December 1996
\end{flushright}
\vspace*{2cm}
\centerline{\Large\bf CP Violation in K Decays}
\centerline{\Large\bf and Rare Decays
\footnote{
Invited Talk presented at the Workshop on Heavy Quarks at Fixed
Target, St. Goar, Germany, 3--6 Oct. 1996, to appear
in the proceedings.
\\
Work supported by the Department of Energy under contract
DE-AC03-76SF00515.}}
\vspace*{1.5cm}
\centerline{{\sc Gerhard Buchalla}}
\bigskip
\centerline{\sl Stanford Linear Accelerator Center}
\centerline{\sl Stanford University, Stanford, CA 94309, U.S.A.}

\vspace*{1.5cm}
\centerline{\bf Abstract}
\vspace*{0.2cm}
\noindent 
The present status of CP violation in decays of neutral kaons
is reviewed. In addition selected rare decays of both $K$ and $B$
mesons are discussed. The emphasis is in particular on observables
that can be reliably calculated and thus offer the possibility of
clean tests of standard model flavor physics.

\vfill

\newpage
\pagenumbering{arabic}

\section{Introduction}
The violation of CP symmetry is one of the most important
issues in contemporary particle physics. First, it is a topic
of fundamental interest in itself.
Together with C violation, CP violation  provides one with an absolute
definition of matter versus antimatter. It is also one of the three
necessary conditions for the generation of a baryon asymmetry
in the universe. 
Studies of this phenomenon allow one furthermore to probe
standard model flavordynamics, which is the part of this theory
that is least understood and contains most of the free model
parameters, including the single CP violating CKM phase $\delta$.
Therefore CP violation is also closely linked to the open question 
of electroweak- and flavor symmetry breaking.
\\
The experimental information on CP violation, on the other hand,
is still very limited. Thirty-two years after the discovery in
$K_L\to\pi^+\pi^-$ decays, its observation has so far been restricted
exclusively to decays of neutral kaons, where it could be identified
in just a handful of modes 
($K_L\to\pi^+\pi^-$, $K_L\to\pi^0\pi^0$, $K_L\to\pi\mu\nu$,
$K_L\to\pi e\nu$ and $K_L\to\pi^+\pi^-\gamma$). All of
these effects are described by a single parameter
$\varepsilon$.
The question of direct CP violation in $K\to\pi\pi$, measured by the
parameter $\varepsilon'/\varepsilon$ is not yet resolved conclusively. 
\\
It is clear from these remarks that further detailed investigations
of CP violation in kaon decays are highly desirable. This includes
$\varepsilon$, $\varepsilon'/\varepsilon$, but also further
possibilities in rare decays such as $K_L\to\pi^0e^+e^-$ or
$K_L\to\pi^0\nu\bar\nu$. Since CP violation and flavordynamics
are intimately related, important additional and complementary
information on this topic will come from studying rare decays in
general, which may or may not be CP violating. Interesting
opportunities are given by $K^+\to\pi^+\nu\bar\nu$ and rare
$B$ decays such as $B\to X_s\gamma$, $B\to X_s\nu\bar\nu$,
$B\to X_sl^+l^-$, $B\to l^+l^-$, and it is natural to include them
in a discussion of CP violation in $K$ decays.
\\
Crucial tests of CP violation will also be conducted by studying
CP asymmetries in decays of $B$ mesons. This very important and
interesting field is covered by the contribution of M. Gronau
(these proceedings) and will therefore not be discussed in this talk.
\\
The outline is as follows. After these introductory remarks we briefly
summarize the theoretical framework that is employed to describe
CP violating and rare decay processes. In section 3 we review the
theoretical status of CP violation in $K\to\pi\pi$ decays, described
by the parameters $\varepsilon$ and $\varepsilon'/\varepsilon$.
The rare decays $K^+\to\pi^+\nu\bar\nu$ and $K_L\to\pi^0\nu\bar\nu$,
the latter of which probes CP violation, are discussed in
section 4. 
Section 5 briefly summarizes the status of $K_L\to\pi^0e^+e^-$ and
$K_L\to\mu^+\mu^-$.
Section 6 addresses the radiative decay
$B\to X_s\gamma$ and the rare decay modes $B\to\mu^+\mu^-$ and
$B\to X_s\nu\bar\nu$ are described in section 7. 
Our emphasis in discussing rare decays 
is on short-distance dominated and theoretically clean
processes, which offer excellent prospects for future precision tests of
SM flavor physics. A selection of further interesting modes is
briefly mentioned in section 8. We conclude with a summary in section 9.
\section{Theoretical Framework}
\begin{figure}[t]
 \vspace{4.5cm}
\includegraphics{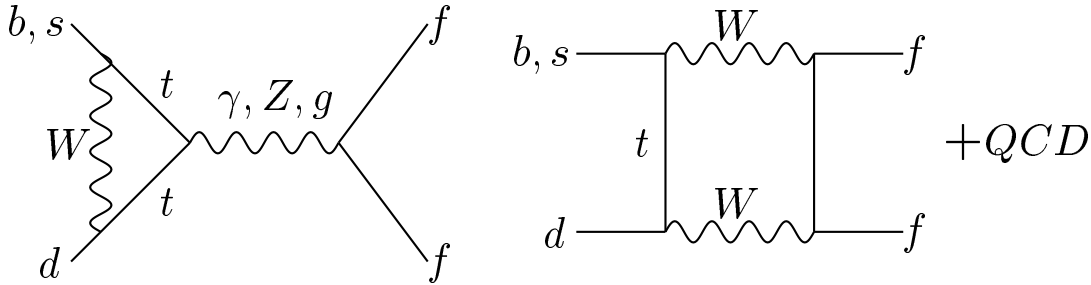}
 \caption{\it
      Typical diagrams for FCNC processes in the standard model.
    \label{fcncfig} }
\end{figure}

In the standard model rare and CP violating decays are related to
loop-induced flavor changing neutral current (FCNC) processes.
This is illustrated in Figure \ref{fcncfig} which shows the
underlying electroweak transitions at the quark level. 
However,
quarks come only in hadronic boundstates. The treatment of FCNC
processes is thus in general a very complex theoretical problem:
It involves electroweak loop transitions at high ($M_W$, $m_t$)
and intermediate ($m_c$) energies in conjunction with QCD radiative
effects at short and long distances, including non-perturbative
strong interaction boundstate dynamics. To make this problem tractable
a systematic approximation scheme is necessary that allows one to
disentangle the interplay of strong and weak interactions.
Such a tool is provided by the operator product expansion (OPE).
It can be used to write the quark level transition in the full
theory, illustrated in Fig. \ref{fcncfig}, in the following form
\begin{equation}\label{heff}
\sum_i\frac{G_F}{\sqrt{2}}V_{CKM}\ C_i(M_W,\mu)\cdot Q_i
\equiv {\cal H}_{eff}
\end{equation}
where the $Q_i$ are a set of local four fermion operators
(usually of dimension six), $C_i$ are the associated Wilson coefficient
functions and $V_{CKM}$ denotes schematically the relevant CKM
parameters. The detailed form and number of the relevant operators
depends on the process under consideration. Operators of dimension higher 
than six are suppressed by inverse powers of the heavy mass scale
(e.g. $M_W$, $m_t$) and can usually be neglected for low energy
$B$ and $K$ meson decays.
\\
Using a somewhat less formal language, operators and Wilson coefficients
are in essence nothing more than effective interaction vertices and
effective couplings, respectively. The expression on the lhs
of (\ref{heff}) can be viewed as a (low energy) effective Hamiltonian,
approximating FCNC interactions among quarks and leptons at energies
far below the $M_W$ scale.
The crucial feature of the OPE approach is that it provides a 
factorization of short distance and long distance contributions.
The short distance physics from scales ${\cal O}(M_W)$ down to
$\mu\;\raisebox{-.4ex}{\rlap{$\sim$}} \raisebox{.4ex}{$>$}\; 1\,GeV$
is factorized into the Wilson coefficients, which can be calculated
in perturbation theory, including QCD effects. The contribution
from long distance scales below $\mu$ on the other hand, is isolated
into the matrix elements of the operators $Q_i$ between physical
hadron states. These have to be treated non-perturbatively,
for instance in lattice QCD (see S. G\"usken, these proceedings).
The scale $\mu$ that separates the short distance and long distance
regime is arbitrary in principle. It has to cancel between the Wilson
coefficients and the matrix elements. For practical purposes, however,
one would like to choose $\mu$ rather low in order to include as much
of the physics as possible into the calculable coefficient function.
On the other hand it is essential for the present approach that QCD
be still perturbative at scale $\mu$, otherwise the calculation of
$C_i$ would break down. Therefore $\mu$ must not be too low. 
Preferably it should also be close to the relevant scale in the
hadronic matrix elements, without of course violating the requirement
of perturbativity. Valid choices are $\mu={\cal O}(m_b)$ for
$B$ decays and
$\mu\;\raisebox{-.4ex}{\rlap{$\sim$}} \raisebox{.4ex}{$>$}\; 1\,GeV$
for $K$ decays.

There has been continuing progress during recent years in our understanding
of FCNC processes within the standard model. First of all, relevant
input parameters have become better known due to ongoing progress in
both theory and experiment.
The most important quantities that enter in constraining the CKM phase 
$\delta$ from the measured value of $\varepsilon$ (kaon CP violation)
are the CKM angles $V_{cb}$, $|V_{ub}/V_{cb}|$, the top quark mass
$m_t$, and the hadronic bag parameter $B_K$.
\\
$V_{cb}$ is already quite well known from exclusive and inclusive
semileptonic $B$ decay, based on heavy quark effective theory (HQET)
\cite{NEU} and heavy quark expansion techniques
\cite{SUV,BBB}, respectively (see also T. Mannel, these proceedings).
For $|V_{ub}|$ the situation is less favorable, but the recent
observation of $B\to(\pi,\,\varrho)l\nu$ at CLEO \cite{CLEO} 
is promising for future improvements on this topic.
The rather precise determination of $m_t$ by CDF and D0 \cite{TIP}
is a remarkable
achievement, in particular for the field of rare decays as it fixes
one of the most important input parameters. Note that the pole mass
value $m^{pole}_t=175\pm 6\,GeV$ measured in experiment corresponds to
a running (${\overline{MS}}$)
mass of $\bar m_t(m_t)=167\pm 6\,GeV$. The latter mass definition
is more suitable for FCNC processes where top appears only as a virtual
particle. The value of $B_K$ from lattice calculations is still not
very precise at present, but systematic uncertainties are becoming
increasingly better under control \cite{SHA,JLQCD}. 
In Table \ref{partab} we summarize the values of the input parameters
that were used for most of the results to be presented below.
The standard model predictions quoted in sections 4, 5 and 7 are
based on \cite{BBL} as updated in \cite{BJLnew}.  
\begin{table}
\centering
\caption{ \it Important input parameters.
}
\vskip 0.1 in
\begin{tabular}{|c|c|c|c|} \hline
$V_{cb}$ & $|V_{ub}/V_{cb}|$ & $\bar m_t(m_t)$ & $B_K$ \\
\hline
$0.040\pm 0.003$ & $0.08\pm 0.02$ & $167\pm 6\,GeV$ & $0.75\pm 0.15$\\
\hline
\end{tabular}
\label{partab}
\end{table}

Further progress has been achieved over the past several years
through the calculation of
next-to-leading order (NLO) QCD corrections in
renormalization group (RG) improved perturbation theory to the
Wilson coefficients for most of the rare and CP violating FCNC
processes. At leading order, leading logarithmic corrections of the
form $(\alpha_s\ln(M_W/\mu))^n$, which are contributions of
${\cal O}(1)$ due to the large logarithm multiplying $\alpha_s$,
are resummed to all orders, $n=0,\,1,\ldots$.
At NLO relative ${\cal O}(\alpha_s)$ corrections of the form
$\alpha_s(\alpha_s\ln(M_W/\mu))^n$ can be systematically included.
This topic is reviewed in \cite{BBL}, where more details and
references can be found. Here we would just like to summarize
the main points that motivate going beyond the leading logarithmic
approximation in weak decay hamiltonians.
\begin{itemize}
\item
First of all, the inclusion of NLO corrections is necessary to test
the validity of perturbation theory.
\item
Without NLO QCD effects a meaningful use of the scheme-specific
QCD scale parameter $\Lambda_{\overline{MS}}$ is not possible.
\item
Unphysical scale dependences can be reduced by going beyond LO.
\item
The Wilson coefficients by themselves are unphysical quantities
and in general scheme dependent. This scheme dependence is an
${\cal O}(\alpha_s)$ (NLO) effect, that is important for a proper
matching to lattice matrix elements.
\item
In some cases the phenomenologically interesting $m_t$-dependence
is, strictly speaking, a NLO effect (e.g. for
$\varepsilon'/\varepsilon$, $K_L\to\pi^0e^+e^-$, $B\to X_se^+e^-$).
\item
If the $m_t$-dependence enters already at leading order
(as is the case e.g. for $K\to\pi\nu\bar\nu$, the top contribution
to $\varepsilon$, $B\to\mu^+\mu^-$ or $B\to X_s\gamma$),
a NLO QCD calculation allows one to make a meaningful distinction
between the running mass $\bar m_t(m_t)\equiv m_t$ and $m^{pole}_t$.
As we have seen the difference of $\approx 8\,GeV$ between both
definitions already exceeds the current experimental error of $6\,GeV$.
\end{itemize}
\section{CP Violation in $K^0\to\pi\pi$ -- 
         $\varepsilon$, $\varepsilon'$}
\subsection{Preliminaries}

CP violation was originally discovered in $K_L\to\pi^+\pi^-$
decays. Among the few cases of CP violation in $K_L$ decays
observed since then, the $\pi\pi$ modes are still the best
studied examples of CP non-conservation and continue to be
under active investigation. The physical neutral kaon states are
$K_L$ and $K_S$ and the two-pion final states they decay to can be
$\pi^+\pi^-$ or $\pi^0\pi^0$. If CP was a good symmetry,
$K_L$ would be CP odd and could not decay into two pions. As a measure
of CP violation one introduces therefore the amplitude ratios
\begin{equation}\label{etapm0}
\eta_{+-}=\frac{\langle\pi^+\pi^-|{\cal T}|K_L\rangle}{
                \langle\pi^+\pi^-|{\cal T}|K_S\rangle}\qquad
\eta_{00}=\frac{\langle\pi^0\pi^0|{\cal T}|K_L\rangle}{
                \langle\pi^0\pi^0|{\cal T}|K_S\rangle}
\end{equation}
If CP violation is entirely due to mixing (indirect CPV),
then $\eta_{+-}=\eta_{00}$. Any difference between $\eta_{+-}$ and
$\eta_{00}$ is thus a measure of direct CP violation. To very good
approximation one may write
\begin{equation}\label{eteps}
\eta_{+-}=\varepsilon+\varepsilon' \qquad
\eta_{00}=\varepsilon-2\varepsilon' 
\end{equation}
where the observable quantities $\varepsilon$ and $\varepsilon'$
parametrize indirect and direct CP violation, respectively.
$\varepsilon'/\varepsilon$ is known to be real up to a phase of a
few degrees. It can thus be measured from the double ratio of rates
\begin{equation}\label{drto}
|\eta_{+-}/\eta_{00}|^2\doteq 1+ 6{\mbox{Re}}\varepsilon'/\varepsilon
\end{equation}
Using standard phase conventions the theoretical expressions for
$\varepsilon$ and $\varepsilon'/\varepsilon$ can be written to very good
approximation as
\begin{equation}\label{eps}
\varepsilon=e^{i\pi/4}\frac{{\mbox{Im}}M_{12}}{\sqrt{2}\Delta M_K}
\end{equation}
\begin{equation}\label{epe}
\frac{\varepsilon'}{\varepsilon}=\frac{\omega}{\sqrt{2}|\varepsilon|}
  \left(\frac{{\mbox{Im}}A_2}{{\mbox{Re}}A_2}-
  \frac{{\mbox{Im}}A_0}{{\mbox{Re}}A_0}
  \right)
\end{equation}
where $M_{12}$ is the off-diagonal element in the
$K^0-\bar K^0$ mass matrix and $\Delta M_K$ the $K_L-K_S$ mass difference.
$A_{0,2}$ are transition amplitudes defined in terms of the strong 
interaction eigenstates $K^0$ and $\pi\pi$ states with definite
isospin ($I=0,\, 2$),
$\langle I=0,2|{\cal T}|K^0\rangle\equiv A_{0,2}\exp(i\delta_{0,2})$.
$\delta_{0,2}$ are strong interaction phases and complexities in 
$A_{0,2}$ arise only from CKM parameters. The smallness of
$\omega\equiv{\mbox{Re}}A_2/{\mbox{Re}}A_0\approx 1/22$ 
reflects the famous $\Delta I=1/2$ rule.
\\
In all current theoretical analyses of $\varepsilon'/\varepsilon$,
the values of $\omega$, $|\varepsilon|$ and ${\mbox{Re}}A_{0,2}$ in
(\ref{epe}) are taken from experiment. ${\mbox{Im}}A_{0,2}$, which
depend on the interesting short-distance physics (top-loops, CKM phase)
are then calculated using an effective Hamiltonian approach (OPE) as
described in section 2.
\\
Experimentally $\varepsilon$ is known very precisely, whereas the
situation with $\varepsilon'/\varepsilon$ is still somewhat unclear.
The current values are
\begin{equation}\label{epsexp}
|\varepsilon|=(2.282\pm 0.019)\cdot 10^{-3}
\end{equation}
\begin{equation}\label{epeexp}
{\mbox{Re}}\frac{\varepsilon'}{\varepsilon}=
\left\{ \begin{array}{ll}
        (23\pm 7)\cdot 10^{-4} & \mbox{NA31\cite{NA31}}\\
        (7.4\pm 5.9)\cdot 10^{-4} & \mbox{E731\cite{E731}}
        \end{array} \right.
\end{equation}

\subsection{Theoretical Status of $\varepsilon$} 

The parameter $\varepsilon$ is determined by the imaginary part
of $M_{12}$ which in turn is generated by the usual
$\Delta S=2$ box-diagrams. The low energy effective Hamiltonian
contains only a single operator $(\bar ds)_{V-A}(\bar ds)_{V-A}$
in this case and one obtains
\begin{equation}\label{epsth}
\varepsilon=e^{i\frac{\pi}{4}}\frac{G^2_F M^2_W f^2_K}{12\pi^2}
\frac{m_K}{\sqrt{2}\Delta M_K}B_K\cdot\mbox{Im}
\left[\lambda^{*2}_c S_0(x_c)\eta_1+\lambda^{*2}_t S_0(x_t)\eta_2
  +2\lambda^*_c\lambda^*_t S_0(x_c,x_t)\eta_3\right]
\end{equation}
Here $\lambda_i=V^*_{is}V_{id}$, $f_K=160\,MeV$ is the kaon decay
constant and the bag parameter $B_K$ is defined by
\begin{equation}\label{bkrsi}
B_K=B_K(\mu)[\alpha^{(3)}_s(\mu)]^{-2/9}
 \left[1+\frac{\alpha^{(3)}_s(\mu)}{4\pi}J_3\right]
\end{equation}
\begin{equation}\label{bkme}
\langle K^0|(\bar ds)_{V-A}(\bar ds)_{V-A}|\bar K^0\rangle
  \equiv\frac{8}{3}B_K(\mu) f^2_K m^2_K
\end{equation}
The index $(3)$ in eq. (\ref{bkrsi}) refers to the number of flavors
in the effective theory and $J_3=307/162$ (in the NDR scheme).
\\
The Wilson coefficient multiplying $B_K$ in (\ref{epsth}) consists
of a charm contribution, a top contribution and a mixed top-charm
contribution. It depends on the quark masses, $x_i\equiv m^2_i/M^2_W$,
through the functions $S_0$. The $\eta_i$ are the corresponding
short-distance QCD correction factors (which depend only slightly on
quark masses). Detailed definitions can be found in \cite{BBL}.
Numerical values for $\eta_1$, $\eta_2$ and $\eta_3$ are summarized
in Table \ref{etaitab}.
\begin{table}
\centering
\caption{ \it NLO results for $\eta_i$ with
$\Lambda^{(4)}_{\overline{MS}}=(325\pm 110)\,MeV$,
$m_c(m_c)=(1.3\pm 0.05)\,GeV$, $m_t(m_t)=(170\pm 15)\,GeV$.
The third column shows the uncertainty due to the errors in
$\Lambda_{\overline{MS}}$ and quark masses. The fourth column
indicates the residual renormalization scale uncertainty at NLO
in the product of $\eta_i$ with the corresponding mass dependent function
from eq. (9). These products are scale independent up to the
order considered in perturbation theory. The central values of the
QCD factors at LO are also given for comparison.
}
\vskip 0.1 in
\begin{tabular}{|c|c|c|c|c|c|} \hline
& NLO(central) & $\Lambda_{\overline{MS}}$, $m_q$ &
scale dep. & NLO ref. & LO(central) \\
\hline
\hline
$\eta_1$ & 1.38 & $\pm 35\%$ & $\pm 15\%$ & \cite{HN1} & 1.12 \\
\hline
$\eta_2$ & 0.574 & $\pm 0.6\%$ & $\pm 0.4\%$ & \cite{BJW} & 0.61 \\
\hline
$\eta_3$ & 0.47 & $\pm 3\%$ & $\pm 7\%$ & \cite{HN3} & 0.35 \\
\hline
\end{tabular}
\label{etaitab}
\end{table}

Concerning these results the following remarks should be made.
\begin{itemize}
\item
$\varepsilon$ is dominated by the top contribution ($\sim 70\%$).
It is therefore rather satisfying that the related short distance
part $\eta_2 S_0(x_t)$ is theoretically extremely well under
control, as can be seen in Table \ref{etaitab}. Note in
particular the very small scale ambiguity at NLO, $\pm 0.4\%$
(for $100\,GeV\leq\mu_t\leq 300\,GeV$). This intrinsic theoretical
uncertainty is much reduced compared to the leading order result
where it would be as large as $\pm 9\%$.
\item
The $\eta_i$ factors and the hadronic matrix element are not
physical quantities by themselves. When quoting numbers it is therefore
essential that mutually consistent definitions are employed.
The factors $\eta_i$ described here are to be used in conjunction
with the so-called scheme- (and scale-) invariant bag parameter
$B_K$ introduced in (\ref{bkrsi}). The last factor on the rhs of
(\ref{bkrsi}) enters only at NLO. As a numerical example, if the
(scale and scheme dependent) parameter $B_K(\mu)$ is given in the
NDR scheme at $\mu=2GeV$, then (\ref{bkrsi}) becomes
$B_K=B_K(NDR,2\,GeV)\cdot 1.31\cdot 1.05$.
\item
The quantity $B_K$ has to be calculated by non-perturbative
methods. Large $N_C$ expansion techniques for instance find
values $B_K=0.75\pm 0.15$ \cite{BBG,GER,BP}. 
The results obtained in other approaches are reviewed in \cite{BBL}.
Ultimately a first
principles calculation should be possible within lattice gauge
theory. Ref. \cite{SHA} quotes an estimate of
$B_K(NDR,2\,GeV)=0.66\pm 0.02\pm 0.11$ in full QCD. The first error
is the uncertainty of the quenched calculation. It is quite small
already and illustrates the progress achieved in controlling
systematic uncertainties in lattice QCD \cite{SHA,JLQCD}. 
The second error represents the
uncertainties in estimating the effects of quenching and 
non-degenerate quark masses.
\end{itemize}

Phenomenologically $\varepsilon$ is used to determine the CKM
phase $\delta$.
\begin{figure}[t]
 \vspace{4.5cm}
\includegraphics{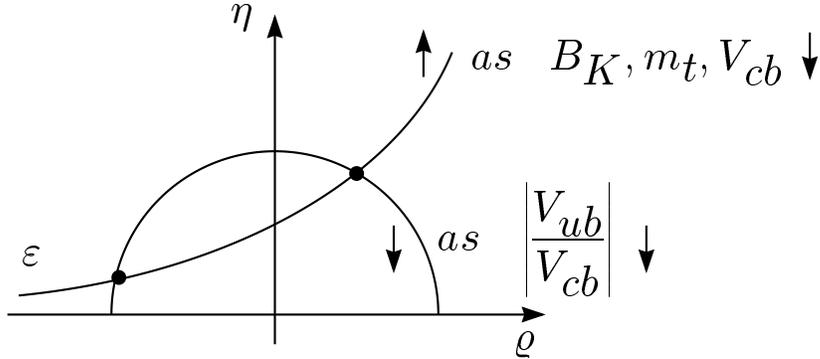}
 \caption{\it
      The $\varepsilon$-constraint on the unitarity triangle.
    \label{utepsfig} }
\end{figure}
The relevant input parameters are
$B_K$, $m_t$, $V_{cb}$ and $|V_{ub}/V_{cb}|$. For fixed
$B_K$, $m_t$ and $V_{cb}$, the measured $|\varepsilon|$ determines
a hyperbola in the $\varrho$--$\eta$ plane of Wolfenstein 
parameters (Figure \ref{utepsfig}).
Intersecting the hyperbola with the circle defined by
$|V_{ub}/V_{cb}|$ determines the unitarity triangle (up to a two-fold
ambiguity). As any one of the four input parameters becomes too small
(with the others held fixed), the SM picture becomes inconsistent.
Using this fact lower bounds on these parameters can be derived
\cite{BUR93}. The large value that has been established for the
top-quark mass in fact helps to maintain the consistency of the SM.

\subsection{Theoretical Status of $\varepsilon'/\varepsilon$}

The expression (\ref{epe}) for $\varepsilon'/\varepsilon$ may
also be written as
\begin{equation}\label{epea02}
\frac{\varepsilon'}{\varepsilon}=-\frac{\omega}{\sqrt{2}
  |\varepsilon|\mbox{Re} A_0}\left(\mbox{Im} A_0-
  \frac{1}{\omega}\mbox{Im} A_2\right)
\end{equation}
$\mbox{Im} A_{0,2}$ are calculated from the general low energy
effective Hamiltonian for $\Delta S=1$ transitions. Including
electroweak penguins this Hamiltonian involves ten different
operators and one has
\begin{equation}\label{ima02}
\mbox{Im} A_{0,2}=-\mbox{Im}\lambda_t \frac{G_F}{\sqrt{2}}
 \sum^{10}_{i=3} y_i(\mu)\langle Q_i\rangle_{0,2}
\end{equation}
Here $y_i$ are Wilson coefficients and
$\langle Q_i\rangle_{0,2}\equiv\langle\pi\pi(I=0,2)|Q_i|K^0\rangle$,
$\lambda_t=V^*_{ts}V_{td}$.
\\
For the purpose of illustration we keep only the numerically
dominant contributions and write
\begin{equation}\label{epeapr}
\frac{\varepsilon'}{\varepsilon}=
\frac{\omega G_F}{2|\varepsilon|\mbox{Re} A_0}\mbox{Im}\lambda_t
\left(y_6\langle Q_6\rangle_0-\frac{1}{\omega}y_8\langle Q_8\rangle_2
+\ldots\right)
\end{equation}
$Q_6$ originates from gluonic penguin diagrams and $Q_8$ from
electroweak contributions. The matrix elements of $Q_6$ and
$Q_8$ can be parametrized by bag parameters $B_6$ and $B_8$ as
\begin{equation}\label{q6me}
\langle Q_6\rangle_0 =-4\sqrt{\frac{3}{2}}
\left[\frac{m_K}{m_s(\mu)+m_d(\mu)}\right]^2 m^2_K(f_K-f_\pi)\cdot B_6
\sim \left(\frac{m_K}{m_s}\right)^2 B_6
\end{equation}
\begin{equation}\label{q8me}
\langle Q_8\rangle_2\simeq\sqrt{3}
\left[\frac{m_K}{m_s(\mu)+m_d(\mu)}\right]^2 m^2_K f_\pi\cdot B_8
\sim \left(\frac{m_K}{m_s}\right)^2 B_8
\end{equation}
$B_6=B_8=1$ corresponds to the factorization assumption for the
matrix elements, which holds in the large $N_C$ limit of QCD.
\\
$y_6\langle Q_6\rangle_0$ and $y_8\langle Q_8\rangle_2$
are positive numbers. The value for $\varepsilon'/\varepsilon$
in (\ref{epeapr}) is thus characterized by a cancellation of
competing contributions. Since the second contribution is an
electroweak effect, suppressed by $\sim\alpha/\alpha_s$ compared
to the leading gluonic penguin $\sim\langle Q_6\rangle_0$,
it could appear at first sight that it should be altogether
negligible for $\varepsilon'/\varepsilon$. However, a number of
circumstances actually conspire to systematically enhance the
electroweak effect so as to render it a very important contribution:
\begin{itemize}
\item
Unlike $Q_6$, which is a pure $\Delta I=1/2$ operator,
$Q_8$ can give rise to the $\pi\pi(I=2)$ final state and thus
yield a non-vanishing $\mbox{Im} A_2$ in the first place.
\item
The ${\cal O}(\alpha/\alpha_s)$ suppression is largely compensated
by the factor $1/\omega\approx 22$ in (\ref{epeapr}), reflecting the
$\Delta I=1/2$ rule.
\item
By contrast to $\langle Q_6\rangle_0$, $\langle Q_8\rangle_2$
is not chirally suppressed ($\langle Q_6\rangle_0$ vanishes in
the chiral limit, where $f_K\to f_\pi$). As a consequence the
matrix element of $Q_8$ is somewhat enhanced relative to the
matrix element of $Q_6$.
\item
$-y_8\langle Q_8\rangle_2$ gives a negative contribution to
$\varepsilon'/\varepsilon$ that strongly grows with $m_t$
\cite{FR,BBH}. For the realistic top mass value it is
quite substantial.
\end{itemize}

The Wilson coefficients $y_i$ have been calculated at NLO
\cite{BJLW,CFMR}. The short-distance part is therefore quite
well under control. The remaining problem is then the computation
of matrix elements, in particular $B_6$ and $B_8$. The cancellation
between these contributions enhances the relative sensitivity of
$\varepsilon'/\varepsilon$ to the anyhow uncertain hadronic parameters
which makes a precise calculation of $\varepsilon'/\varepsilon$
impossible at present. The results found in various recent analyses
are collected in Table \ref{epetab}.
\begin{table}
\centering
\caption{ \it Estimates of $B_6$ and $B_8$ and calculations of
$\varepsilon'/\varepsilon$. (g) refers to the assumption of a
Gaussian distribution of errors in the input parameters,
(s) to the more conservative 'scanning' of parameters over their
full allowed ranges.
}
\vskip 0.1 in
\begin{tabular}{|c|c|c|c|c|}\hline
$B_6$ & $B_8$ & $B_{6,8}$ ref. & $\varepsilon'/\varepsilon$ ref. &
  $(\varepsilon'/\varepsilon)/10^{-4}$ \\
\hline
\hline
$1.0\pm 0.2$ & $1.0\pm 0.2$ & large $N_C$ \cite{BG} &
 \cite{BJL96} & $[-1.2, 16]$ (s) \\
 & & & & $[0.2, 7.0]$ (g) \\
\hline
$1.0\pm 0.2$ & $1.0\pm 0.2$ & lattice \cite{FMMM,BS,KIL,SHA2} &
 \cite{CFMRS} & $[0.6, 5.6]$ (g) \\
\hline
$1.0\pm 0.4$ & $2.2\pm 1.5$ & chiral quark model \cite{BEF1} &
 \cite{BEF2} & $[-50, 14]$ (s) \\
\hline
$\sim 1.3$ & $\sim 0.7$ & \cite{FRO} &
 \cite{HEI} & $[5.8, 14.0]$  \\
\hline
\end{tabular}
\label{epetab}
\end{table}

Recently, the issue of the strange quark mass has received increased 
attention due to new lattice results reporting lower than
anticipated values.
As we have seen in (\ref{q6me}), (\ref{q8me}) the matrix elements
of $Q_6$ and $Q_8$ are expected to behave as $1/m^2_s$, up to 
$B$-factors. This result is based on the factorization ansatz,
which holds in the large $N_C$ limit of QCD, and reflects the 
particular, scalar-current type structure of $Q_6$ and $Q_8$.
The phenomenological predictions thus show a marked dependence
on the strange quark mass used in the analysis. Generally
$\varepsilon'/\varepsilon$ will increase with decreasing $m_s$.
The estimates for $\varepsilon'/\varepsilon$ in Table \ref{epetab}
are based on strange quark masses in the ball park of
$m_s(2GeV)=130\,MeV$. Table \ref{mstab} collects a few recent
determinations of $m_s$ from QCD sum rules and from lattice
calculations.
\begin{table}
\centering
\caption{ \it Results for the running strange quark mass
(in the ${\overline{MS}}$ scheme). The lattice results correspond to
the quenched approximation. The numbers in brackets are
estimates for the unquenched case.
}
\vskip 0.1 in
\begin{tabular}{|c|c|} \hline
\multicolumn{2}{|l|}{$m_s(2\, GeV)/MeV$} \\
\hline
\hline
$145\pm 20$ & QCD sum rules \cite{JM,CDPS,NAR1}\\
\hline
$127\pm 18$ & lattice (Rome) \cite{ACCLM}\\
\hline
$90\pm 20$ ($55-70$) & lattice (Los Alamos) \cite{GBH}\\
\hline
$95\pm 16$ ($54-92$) & lattice (Fermilab) \cite{GOU} \\
\hline
\end{tabular}
\label{mstab}
\end{table}

Using the low $m_s$ values indicated by the very recent
Los Alamos and Fermilab lattice results Buras et al. \cite{BJL96}
find
\begin{equation}
0\leq\varepsilon'/\varepsilon\leq 43\cdot 10^{-4} \mbox{(scanning)}
\end{equation}
\begin{equation}
2.1\cdot 10^{-4}\leq\varepsilon'/\varepsilon\leq 18.7\cdot 10^{-4} 
\mbox{(Gaussian)}
\end{equation}
for $m_s(2\, GeV)=(86\pm 17)\, MeV$.
This is compatible with both experimental results (\ref{epeexp}),
within the rather large uncertainties. Using $m_s(2\, GeV)$ around
$130\, MeV$, on the other hand, the results are consistent with E731,
but somewhat low compared to NA31 (see the first line of 
Table \ref{epetab}).

In conclusion, the SM prediction for $\varepsilon'/\varepsilon$
suffers from large hadronic uncertainties, reinforced by substantial
cancellations between the $I=0$ and $I=2$ contributions. Despite this
problem, the characteristic pattern of CP violation observed in
$K\to\pi\pi$ decays, namely $\varepsilon={\cal O}(10^{-3})$ and
$\varepsilon'={\cal O}(10^{-6})$ (or below), is well accounted for
by the standard theory, which can be considered a non-trivial success
of the model.
\\
On the experimental side a clarification of the current situation is to 
be expected by the upcoming new round of $\varepsilon'/\varepsilon$
experiments conducted at Fermilab (E832), CERN (NA48) and
Frascati (KLOE). The goal is a measurement of $\varepsilon'/\varepsilon$
at the $10^{-4}$ level. The demonstration that $\varepsilon'\not= 0$
would constitute a qualitatively new feature of CP violation and
as such be of great importance. 
However, due to the large uncertainties in the
theoretical calculation, a quantitative use of this result for the
extraction of CKM parameters will unfortunately be severely limited.
For this purpose one has to turn to theoretically cleaner
observables. As we will see in the next section, rare
$K$ decays in fact offer very promising opportunities in this direction.

\section{The Rare Decays $K^+\to\pi^+\nu\bar\nu$ and
$K_L\to\pi^0\nu\bar\nu$}

\begin{figure}[t]
 \vspace{5cm}
\includegraphics{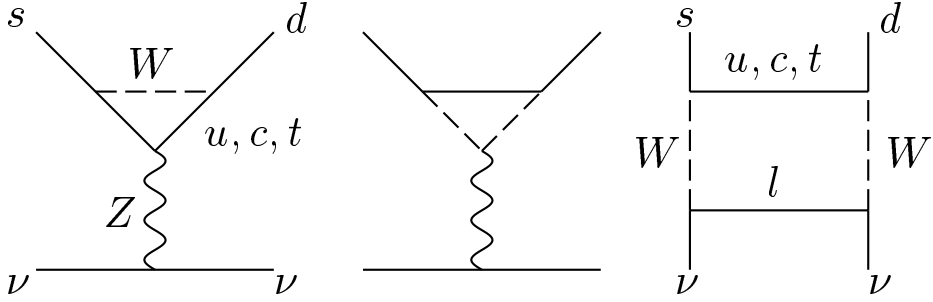}
 \caption{\it
      Leading order electroweak diagrams contributing to
      $K\to\pi\nu\bar\nu$ in the standard model.
    \label{kpnnfig} }
\end{figure}

The decays $K\to\pi\nu\bar\nu$ proceed through flavor changing
neutral current effects. These arise in the standard model only at
second (one-loop) order in the electroweak interaction
(Z-penguin and W-box diagrams, Figure \ref{kpnnfig}) 
and are additionally GIM suppressed.
The branching fractions are thus very small, at the level of
$10^{-10}$, which makes these modes rather challenging to detect.
However, $K\to\pi\nu\bar\nu$ have long been known to be reliably
calculable, in contrast to most other decay modes of interest.
A measurement of $K^+\to\pi^+\nu\bar\nu$ and $K_L\to\pi^0\nu\bar\nu$
will therefore be an extremely useful test of flavor physics.
Over the recent years important refinements have been added to the
theoretical treatment of $K\to\pi\nu\bar\nu$. These have helped to
precisely quantify the meaning of the term 'clean' in this context
and have reinforced the unique potential of these observables.
Let us briefly summarize the main aspects of why $K\to\pi\nu\bar\nu$
is theoretically so favorable and what recent developments have
contributed to emphasize this point.
\begin{itemize}
\item
First, $K\to\pi\nu\bar\nu$ is semileptonic. The relevant hadronic
matrix elements $\langle\pi|(\bar sd)_{V-A}|K\rangle$ are just
matrix elements of a current operator between hadronic states, which are
already considerably simpler objects than the matrix elements of
four-quark operators encountered in many other observables
($K-\bar K$ mixing, $\varepsilon'/\varepsilon$).
Moreover, they are related to the matrix element
\begin{equation}\label{sume}
\langle\pi^0|(\bar su)_{V-A}|K^+\rangle
\end{equation} 
by isospin symmetry.
The latter quantity can be extracted from the well measured leading
semileptonic decay $K^+\to\pi^0 l\nu$. Although isospin is a fairly
good symmetry, it is still broken by the small up- down quark mass
difference and by electromagnetic effects. These manifest themselves
in differences of the neutral versus charged kaon (pion) masses
(affecting phase space), corrections to the isospin limit in the
formfactors and electromagnetic radiative effects.
Marciano and Parsa \cite{MP} have analyzed these corrections and
found an overall reduction in the branching ratio by $10\%$ for
$K^+\to\pi^+\nu\bar\nu$ and $5.6\%$ for $K_L\to\pi^0\nu\bar\nu$.
\item
Long distance contributions are systematically suppressed 
as ${\cal O}(\Lambda^2_{QCD}/m^2_c)$ compared to the 
charm contribution (which is part of the short distance amplitude).
This feature is related to the hard ($\sim m^2_c$) GIM suppression
pattern shown by the Z-penguin and W-box diagrams, and the absence
of virtual photon amplitudes. Long distance contributions have
been examined quantitatively \cite{RS,HL,LW,GHL,FAJ} and shown to be
numerically negligible (below $\approx 5\%$ of the charm amplitude).
\item
The preceeding discussion implies that $K\to\pi\nu\bar\nu$ are
short distance dominated (by top- and charm-loops in general).
The relevant short distance QCD effects can be treated in
perturbation theory and have been calculated at next-to-leading
order \cite{BB2,BB3}. This allowed to substantially reduce
(for $K^+$) or even practically eliminate ($K_L$) the leading
order scale ambiguities, which are the dominant uncertainties in
the leading order result.
\end{itemize}

In Table \ref{kpnntab} we have summarized some of the main
features of $K^+\to\pi^+\nu\bar\nu$ and $K_L\to\pi^0\nu\bar\nu$.
\begin{table}
\centering
\caption{ \it Compilation of important properties and results
for $K\to\pi\nu\bar\nu$.
}
\vskip 0.1 in
\begin{tabular}{|c|c|c|} \hline
& $K^+\to\pi^+\nu\bar\nu$ & $K_L\to\pi^0\nu\bar\nu$ \\
\hline
\hline
& CP conserving & CP violating \\
\hline
CKM & $V_{td}$ & $\mbox{Im} V^*_{ts}V_{td}\sim J_{CP}\sim\eta$ \\
\hline
contributions & top and charm & only top \\
\hline
scale uncert. (BR) & $\pm 20\%$ (LO) $\to \pm 5\%$ (NLO) &
                     $\pm 10\%$ (LO) $\to \pm 1\%$ (NLO) \\
\hline
BR (SM) & $(0.9\pm 0.3)\cdot 10^{-10}$&$(2.8\pm 1.7)\cdot 10^{-11}$ \\
\hline
exp. limit & $< 2.4\cdot 10^{-9}$ BNL 787 \cite{ADL}
           & $< 5.8\cdot 10^{-5}$ FNAL 799 \cite{WEA} \\
\hline
\end{tabular}
\label{kpnntab}
\end{table}
The neutral mode proceeds through CP violation in the standard model.
This is due to the definite CP properties of $K^0$, $\pi^0$ and
the hadronic transition current $(\bar sd)_{V-A}$. The violation of
CP symmetry in $K_L\to\pi^0\nu\bar\nu$ arises through interference
between $K^0-\bar K^0$ mixing and the decay amplitude. This mechanism
is sometimes refered to as mixing-induced CP violation. Now, in the
standard model, the mixing-induced CP violation in $K_L\to\pi^0\nu\bar\nu$
is by orders of magnitude larger than the one in $K_L\to\pi^+\pi^-$,
for instance. Any difference in the magnitude of mixing induced
CP violation between two $K_L$ decay modes is a signal of direct
CP violation. 
In this sense, the standard model decay $K_L\to\pi^0\nu\bar\nu$ is
a signal of almost pure direct CP violation, revealing an effect
that can not be explained by CP violation in the $K-\bar K$
mass matrix alone.
\\
While already $K^+\to\pi^+\nu\bar\nu$ can be reliably calculated,
the situation is even better for $K_L\to\pi^0\nu\bar\nu$. Since
only the imaginary part of the amplitude (in standard phase
conventions) contributes, the charm sector, in $K^+\to\pi^+\nu\bar\nu$
the dominant source of uncertainty, is completely negligible for
$K_L\to\pi^0\nu\bar\nu$ ($0.1\%$ effect on the branching ratio).
Long distance contributions 
($\;\raisebox{-.4ex}{\rlap{$\sim$}} \raisebox{.4ex}{$<$}\; 0.1\%$)  
and also the indirect CP violation effect  
($\;\raisebox{-.4ex}{\rlap{$\sim$}} \raisebox{.4ex}{$<$}\; 1\%$)  
are likewise negligible. In summary, the total theoretical
uncertainties, from perturbation theory in the top sector
and in the isospin breaking corrections, are safely below
$2-3\%$ for $B(K_L\to\pi^0\nu\bar\nu)$. This makes this decay
mode truly unique and very promising for phenomenological
applications. (Note that the range given as the standard model
prediction in Table \ref{kpnntab} arises from our, at present,
limited knowledge of standard model parameters (CKM), and not
from intrinsic uncertainties in calculating $B(K_L\to\pi^0\nu\bar\nu)$).

\begin{figure}[t]
 \vspace{5cm}
\includegraphics{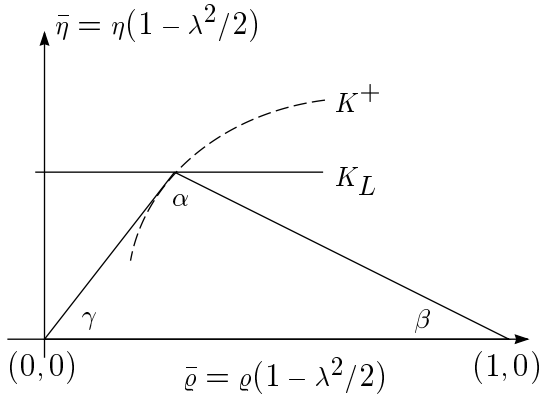}
 \caption{\it
      Unitarity triangle from $K\to\pi\nu\bar\nu$.
    \label{utkpnfig} }
\end{figure}

With a measurement of $B(K^+\to\pi^+\nu\bar\nu)$ and 
$B(K_L\to\pi^0\nu\bar\nu)$ available very interesting phenomenological
studies could be performed. 
For instance, $B(K^+\to\pi^+\nu\bar\nu)$ and 
$B(K_L\to\pi^0\nu\bar\nu)$ together determine the unitarity triangle
(Wolfenstein parameters $\varrho$ and $\eta$) 
completely (Figure \ref{utkpnfig}). 
The expected accuracy with $\pm 10\%$ branching ratio measurements is
comparable to the one that can be achieved by CP violation studies
at $B$ factories before the LHC era \cite{BB96}.
The quantity $B(K_L\to\pi^0\nu\bar\nu)$ by itself offers probably the
best precision in determining $\mbox{Im} V^*_{ts}V_{td}$ or,
equivalently, the Jarlskog parameter
\begin{equation}\label{jcp}
J_{CP}=\mbox{Im}(V^*_{ts}V_{td}V_{us}V^*_{ud})=
\lambda\left(1-\frac{\lambda^2}{2}\right)\mbox{Im}\lambda_t
\end{equation}
The prospects here are even better than for $B$ physics at the LHC.
As an example, let us assume the following results will be
available from B physics experiments
\begin{equation}\label{lhcb}
\sin 2\alpha=0.40\pm 0.04\quad \sin 2\beta=0.70\pm 0.02\quad
V_{cb}=0.040\pm 0.002
\end{equation}
The small errors quoted for $\sin 2\alpha$ and $\sin 2\beta$ from
CP violation in $B$ decays require precision measurements at the LHC.
In the case of $\sin 2\alpha$ we have to assume in addition that the
theoretical problem of 'penguin-contamination' can be resolved.
These results would then imply
$\mbox{Im}\lambda_t=(1.37\pm 0.14)\cdot 10^{-4}$.
On the other hand, a $\pm 10\%$ measurement 
$B(K_L\to\pi^0\nu\bar\nu)=(3.0\pm 0.3)\cdot 10^{-11}$ together with
$m_t(m_t)=(170\pm 3) GeV$ would give
$\mbox{Im}\lambda_t=(1.37\pm 0.07)\cdot 10^{-4}$. If we are optimistic
and take $B(K_L\to\pi^0\nu\bar\nu)=(3.0\pm 0.15)\cdot 10^{-11}$,
$m_t(m_t)=(170\pm 1) GeV$, we get
$\mbox{Im}\lambda_t=(1.37\pm 0.04)\cdot 10^{-4}$, a truly remarkable
accuracy. The prospects for precision tests of the standard model
flavor sector will be correspondingly good.

The charged mode $K^+\to\pi^+\nu\bar\nu$ is being currently
pursued by Brookhaven experiment E787. The latest published result
\cite{ADL} gives an upper limit which is about a factor 25 above
the standard model range. Several improvements have been implemented
since then and the SM sensitivity is expected to be reached in the
near future \cite{AGS2}. For details see the contribution of
S. Kettell (these proceedings).
Recently an experiment has been proposed to measure 
$K^+\to\pi^+\nu\bar\nu$ at the Fermilab Main Injector \cite{CCTR}.
Concerning $K_L\to\pi^0\nu\bar\nu$, a proposal exists at
Brookhaven (BNL E926) to measure this decay at the AGS 
with a sensitivity of ${\cal O}(10^{-12})$ (see \cite{AGS2}).
There are furthermore plans to pursue this mode with comparable
sensitivity at Fermilab \cite{KAMI} and KEK \cite{ISS}.
It will be very exciting to follow the development and outcome of
these ambitious projects. The 'holy grail of kaon physics' could
finally come within reach.

\section{$K_L\to\pi^0e^+e^-$ and $K_L\to\mu^+\mu^-$}

\subsection{$K_L\to\pi^0e^+e^-$}

This decay mode has obvious similarities with $K_L\to\pi^0\nu\bar\nu$
and the apparent experimental advantage of charged leptons, rather than
neutrinos, in the final state. However there are a number of quite
serious difficulties associated with this very fact. Unlike neutrinos the
electron couples to photons. As a consequence the amplitude, which 
was essentially purely short distance in $K_L\to\pi^0\nu\bar\nu$, becomes
sensitive to poorly calculable long distance physics (photon penguin).
Simultaneously the importance of indirect CP violation 
($\sim\varepsilon$) is strongly enhanced and furthermore a long distance
dominated, CP conserving amplitude with two-photon intermediate state
can contribute significantly. Treating $K_L\to\pi^0e^+e^-$
theoretically one is thus faced with the need to disentangle three
different contributions of roughly the same order of magnitude.
\begin{itemize}
\item
Direct CP violation: This part is short distance in character,
theoretically clean and has been analyzed at next-to-leading order
in QCD \cite{BLMM}. Taken by itself this mechanism leads to a
$K_L\to\pi^0e^+e^-$ branching ratio of $(4.5\pm 2.6)\cdot 10^{-12}$
within the standard model.
\item
Indirect CP violation: This amplitude is determined through
$\sim\varepsilon\cdot A(K_S\to\pi^0e^+e^-)$. The $K_S$ amplitude is
dominated by long distance physics and has been investigated in
chiral perturbation theory \cite{EPR,BRP,DG}. Due to unknown counterterms
that enter this analysis a reliable prediction is not possible at
present. The situation would improve with a measurement of
$B(K_S\to\pi^0e^+e^-)$, which could become possible at DA$\Phi$NE.
Present day estimates for $B(K_L\to\pi^0e^+e^-)$ due to indirect
CP violation alone give typically values of
$(1-5)\cdot 10^{-12}$.
\item
The CP conserving two-photon contribution is again long-distance
dominated. It has been analyzed by various authors \cite{DG,CEP,HS}.
The estimates are typically a few $10^{-12}$. Improvements in this sector
might be possible by further studying the related decay
$K_L\to\pi^0\gamma\gamma$ whose branching ratio has already been
measured to be $(1.7\pm 0.3)\cdot 10^{-6}$. 
\end{itemize}

Originally it had been hoped for that the direct CP violating
contribution is dominant. Unfortunately this could so far not be
unambiguously established and requires further study.
\\
Besides the theoretical problems, $K_L\to\pi^0e^+e^-$ is also very hard
from an experimental point of view. The expected branching ratio 
is even smaller than for $K_L\to\pi^0\nu\bar\nu$. Furthermore a
serious irreducible physics background from the radiative mode
$K_L\to e^+e^-\gamma\gamma$ has been identified, which poses additional
difficulties \cite{LV}. A background subtraction seems necessary,
which is possible with enough events. 
Additional information could in principle also be gained by studying
the electron energy asymmetry \cite{DG,HS} or the time evolution
\cite{DG,LIT,KP}.

\subsection{$K_L\to\mu^+\mu^-$}

$K_L\to\mu^+\mu^-$ receives a short distance contribution from
Z-penguin and W-box graphs similar to $K\to\pi\nu\bar\nu$. This
part of the amplitude is sensitive to the Wolfenstein parameter
$\varrho$. In addition $K_L\to\mu^+\mu^-$ proceeds through a
long distance contribution with two-photon intermediate state,
which actually dominates the decay completely. The long distance
amplitude consists of a dispersive ($A_{dis}$) and an absorbtive
contribution ($A_{abs}$). The branching fraction can thus be written
\begin{equation}\label{bklma}
B(K_L\to\mu^+\mu^-)=|A_{SD}+A_{dis}|^2 + |A_{abs}|^2
\end{equation}
Using $B(K_L\to\gamma\gamma)$ it is possible to extract
$|A_{abs}|^2=(6.8\pm 0.3)\cdot 10^{-9}$ \cite{LV}.
$A_{dis}$ on the other hand can not be calculated accurately at 
present and the estimates are strongly model dependent 
\cite{BMS,GN,BEG,KO,EKP}.
This is rather unfortunate, in particular since 
$B(K_L\to\mu^+\mu^-)$, unlike most other rare decays, has already
been measured, and this with very good precision
\begin{equation}\label{bklmex}
B(K_L\to\mu^+\mu^-)=
\left\{ \begin{array}{ll}
        (6.9\pm 0.4)\cdot 10^{-9} & \mbox{BNL 791 \cite{HEIN}}\\
        (7.9\pm 0.7)\cdot 10^{-9} & \mbox{KEK 137 \cite{AKA}}
        \end{array} \right.
\end{equation}
For comparison we note that 
$B(K_L\to\mu^+\mu^-)_{SD}=(1.3\pm 0.6)\cdot 10^{-9}$ is the
expected branching ratio in the standard model based on the
short-distance contribution alone. Due to the fact that $A_{dis}$
is largely unknown, $K_L\to\mu^+\mu^-$ is at present not a very
useful constraint on CKM parameters.

\section{The Radiative Rare Decay $B\to X_s\gamma$}

The radiative decay $B\to X_s\gamma$ is justifiably one of the
highlights in the field of flavor changing neutral currents.
First of all, its rate is of order $G^2_F\alpha$, while most other
FCNC processes are only $\sim G^2_F\alpha^2$. This leads to a relatively
sizable branching fraction of ${\cal O}(10^{-4})$. The decay is
accessible to experiment already today and its branching ratio
has been measured at CLEO \cite{ALA}
\begin{equation}\label{bsgex}
B(B\to\ X_s\gamma)=(2.32\pm 0.67)\cdot 10^{-4}
\end{equation}
At the same time the inclusive transition $B\to X_s\gamma$ can be
systematically treated by standard theoretical techniques such as
heavy quark expansion and renormalization group improved perturbation
theory. It is sensitive to short distance physics and provides 
therefore a good test of flavordynamics. Extensions of the standard
model, for instance the Two-Higgs-Doublet Model 
\cite{GSW,HEW1,BBP,HMT,BMMP},
models with three Higgs-doublets \cite{GRN},
minimal SUSY \cite{BBMR,BAG,BOR} or left-right symmetric models \cite{CM}
receive important constraints from $B\to X_s\gamma$.
\\
In the following we shall briefly sketch the theoretical status
of $B\to X_s\gamma$ in the standard model.
\\
The basic structure of the $b\to s\gamma$ transition is quite interesting
from a theoretical point of view. Schematically one has, in the
leading log approximation \cite{CFRS}
\begin{equation}\label{bsgam}
B(B\to X_s\gamma)\sim\left|F(m_t)+\sum_n \sim\left(
  \alpha_s \ln\frac{M_W}{\mu}\right)^n\right|^2
\end{equation}
$F(m_t)$ is a function describing the top quark mass dependence.
The large logarithmic QCD corrections $\sim\alpha_s\ln(M_W/\mu)$,
$\mu={\cal O}(m_b)$, are resummed to all orders. Their contribution
is formally ${\cal O}(1)$, of the same order as $F(m_t)$.
Technically these effects, although of leading order, are generated
from two-loop contributions, whereas usually leading logarithmic
effects arise at the one-loop level. This peculiarity is due to the 
radiative nature of the FCNC in $b\to s\gamma$.
Numerically one finds $B(B\to X_s\gamma)\approx 1.2\cdot 10^{-4}$
neglecting all QCD effects, but $\approx 2.8\cdot 10^{-4}$ including
the tower of leading logarithmic corrections. This illustrates the
decisive impact of short distance QCD effects on the prediction of
$B(B\to X_s\gamma)$. With this feature $B\to X_s\gamma$ is the
prototype example for the importance of perturbative QCD corrections
in weak decays.
\\
A somewhat unwelcome side effect of the predominance of QCD
contributions is the rather strong scale ($\mu$) ambiguity of the result 
at leading order \cite{AG,BMMP}, 
implying an uncertainty of $\pm 25\%$ in the
braching fraction (for $m_b/2\leq\mu\leq m_b$). This is the dominant
uncertainty in the leading order prediction of $B(B\to X_s\gamma)$.
Several other, somewhat less prominent sources of error exist.
\begin{itemize}
\item
Long distance contributions arise from intermediate $(c\bar c)$
bound states coupling to the on-shell photon. Their impact on the
branching ratio is expected to be of the order
$\;\raisebox{-.4ex}{\rlap{$\sim$}} \raisebox{.4ex}{$<$}\; 10\%$
\cite{DHT,SOA,EIMS}.
\item
The theoretical prediction of $B(B\to X_s\gamma)$ is normalized
to $B(B\to X_c l\nu)$, which depends on $m_c/m_b$. The corresponding 
error is about $6\%$.
\item
The ratio $|V^*_{ts}V_{tb}/V_{cb}|^2$
entering $B(B\to X_s\gamma)/B(B\to X_c l\nu)$ 
is quite well constrained to $0.95\pm 0.03$ from CKM unitarity and
using input from $\varepsilon_K$ and $B-\bar B$ mixing.
\item
The uncertainty from the error in $\alpha_s(M_Z)$ is 
$\;\raisebox{-.4ex}{\rlap{$\sim$}} \raisebox{.4ex}{$<$}\; 10\%$.
The errors due to the experimental values for $B(B\to X_c l\nu)$
and $m_t$ are small.
\item
Non-perturbative contributions to $B(B\to X_s\gamma)$ from
subleading terms $(\sim 1/m^2_b)$ in the heavy quark expansion
have also been analyzed \cite{BBSUV}. They are likewise negligible.
\end{itemize}

The essential step for further improvement is therefore a complete
and consistent NLO calculation. 
For an overview of the various parts of such an analysis and
detailed references see \cite{BBL}.
The last two major ingredients
in this very complex calculation have recently been performed and
results were reported at the 28th International Conference on
High Energy Physics (ICHEP 96) in Warsaw. NLO QCD corrections to
the matrix elements have been addressed by Greub, Hurth and Wyler
\cite{GHW} and the three-loop contribution to the NLO renormalization
group evolution has been worked out by Chetyrkin, Misiak and M{\"u}nz
\cite{CMM}.
The {\em preliminary} result reads
\begin{equation}\label{bsgnlo}
B(B\to X_s\gamma)=(3.3\pm 0.5)\cdot 10^{-4}\qquad\mbox{(NLO, preliminary)}
\end{equation}
The error represents the total uncertainty, including the one from 
residual scale dependence. The latter has decreased as expected,
from $\pm 25\%$ to about $\pm 6\%$ after incorporating the NLO
corrections.
Eq. (\ref{bsgnlo}) can be compared with the leading order result
$B(B\to X_s\gamma)_{LO}=(2.8\pm 0.8)\cdot 10^{-4}$ and with the
experimental number in (\ref{bsgex}). Although the central value
of (\ref{bsgnlo}) is apparently higher than the experimental
$2.32\cdot 10^{-4}$, it is still premature to draw definitive
conclusions.
\\
Exclusive channels, such as $B\to K^*\gamma$, have also been
studied \cite{BAL,ABS,NAR2,FGM}, but are more difficult from
a theoretical point of view.

\section{The Rare Decays $B_s\to\mu^+\mu^-$ and
$B\to X_s\nu\bar\nu$}
These decays are both theoretically very clean since they
are entirely dominated by virtual top contributions which
proceed at very short distances. The relevant Feynman graphs are
Z-penguin and W-box diagrams similar to those for $K\to\pi\nu\bar\nu$.
Next-to-leading order QCD corrections essentially eliminate the
leading order scale uncertainty of $\pm 10\%$ to merely $\pm 1\%$
in the branching ratios \cite{BB2}. 

The branching ratio for $B_s\to\mu^+\mu^-$ is proportional to
$|V_{ts}|^2$ and $f^2_{B_s}$. Detailed expressions can be found
in \cite{BBL}. The standard model expectation is
$B(B_s\to\mu^+\mu^-)=(3.6\pm 1.8)\cdot 10^{-9}$, based on
$f_{B_s}=(210\pm 30)\,MeV$. The current experimental upper
limit on the branching ratio is $8.4\cdot 10^{-6}$ \cite{KRO}.
\\
For the related mode $B_d\to\mu^+\mu^-$ the theoretical prediction
is about an order of magnitude lower than for $B_s\to\mu^+\mu^-$
and an upper limit of $1.6\cdot 10^{-6}$ has been set by CDF
\cite{KRO}. The decays could become accessible at the LHC.
Their ratio
\begin{equation}\label{bdbsmm}
\frac{B(B_d\to\mu^+\mu^-)}{B(B_s\to\mu^+\mu^-)}=
\frac{\tau(B_d)}{\tau(B_s)}\frac{m_{B_d}}{m_{B_s}}\cdot
\frac{f^2_{B_d}}{f^2_{B_s}} \left|\frac{V_{td}}{V_{ts}}\right|^2
\end{equation}
is a measure of $|V_{td}/V_{ts}|$, once $SU(3)$ breaking effects
in $f_{B_d}/f_{B_s}$ are properly taken into account.
Results for other final states, $B_{d,s}\to e^+e^-$ or
$\tau^+\tau^-$ are summarized in \cite{BBL}.

The inclusive decay $B\to X_s\nu\bar\nu$ is similar to 
$B_s\to\mu^+\mu^-$.
The disadvantage is a more challenging experimental signature.
Advantages of $B\to X_s\nu\bar\nu$ over $B_s\to\mu^+\mu^-$,
on the other hand, are the absence of the strong helicity suppression,
resulting in a much larger branching fraction, and the inclusive
nature of the decay, which allows a reliable calculation of the
matrix element with heavy quark expansion and perturbative QCD.
The ratio of 
$B(B\to X_s\nu\bar\nu)/B(B\to X_d\nu\bar\nu)$ is a clean measure
of $|V_{td}/V_{ts}|$.
\\
The decay $B\to X_s\nu\bar\nu$ received renewed interest after a
proposal to extract an upper limit on its branching fraction from
available data by Grossman et al. \cite{GLN}. Subsequently this led
to an upper bound of $7.7\cdot 10^{-4}$ by the ALEPH collaboration
\cite{ALEPH}, already quite close to the standard model range
$B(B\to X_s\nu\bar\nu)=(3.8\pm 0.8)\cdot 10^{-5}$. The result
constrains scenarios of new physics \cite{GLN}. In view of
the experimental situation and the theoretically clean character
$B\to X_s\nu\bar\nu$ clearly deserves further attention.

\section{Other Opportunities}

There are several other possibilities to investigate flavor physics
by studying rare decay modes.

In the field of $B$ decays the inclusive mode $B\to X_sl^+l^-$
($l=e,\mu,\tau$), for instance, has been widely discussed in the
literature. The next-to-leading order QCD corrections are known
\cite{BM,MIS}. The decay branching ratio, dilepton invariant mass spectrum,
forward-backward charge asymmetry and lepton polarization could be 
useful probes of
the standard model and its extensions \cite{AGM,HEW2,KS,CMW,HW}.

A particular class of rare kaon decays are the modes
$K_L\to\mu e$, $K^+\to\pi^+\mu e$ and $K_L\to\pi^0\mu e$, which
violate lepton flavor and are altogether forbidden in the standard model.
Current limits for their branching ratios are at the level of
$\sim 10^{-10}$. They will be improved by future experiments 
(see the talks by S. Pislak, W. Molzon and E. Ramberg, these proceedings)
down to the $10^{-12}$ level, corresponding to a sensitivity to
scales of typically a few hundred $TeV$. This might be a way to
probe, albeit indirectly, high energy scales not accessible by any 
other method.

Besides $K$ and $B$ physics, also $D$ mesons might yield
interesting clues on flavordynamics. Here standard model effects are
generally very small and long-distance contributions usually play an
essential role. Still the charmed meson sector could provide a window
for new physics. This topic has been reviewed by Burdman \cite{BURD},
where more details and references can be found.
A general reference for new physics in FCNC processes is
Hewett et al. \cite{HTT}.

\section{Summary}

We have reviewed the present status of CP violation in kaon
decays and discussed selected rare decays of both $K$ and $B$ mesons.
To conclude we summarize some of the main issues.
\begin{itemize}
\item
The field of CP violation and rare decays is an important
probe of flavordynamics.
\item
Short distance QCD corrections have by now been calculated at
next-to-leading order for almost all cases of practical interest.
\item
So far the parameter $\varepsilon$ in the neutral kaon system is
still the only signal of CP violation observed in the laboratory.
Important phenomenological constraints can be derived from this
measurement.
\item
The situation of whether $\varepsilon'/\varepsilon$ is zero or not
will soon be clarified experimentally with an accuracy of $10^{-4}$.
This could establish an important, qualitatively new aspect of
CP violation. The quantitative use of this result for the extraction
of CKM parameters, however, is severely limited by large hadronic
uncertainties.
\item
Precise extractions of CKM quantities along with accurate standard
model tests will be possible with theoretically clean observables.
A prime candidate is the 'golden reaction' $K_L\to\pi^0\nu\bar\nu$,
which is in particular an ideal measure of the Jarlskog parameter
$J_{CP}$.
\item
Complemetary information from as many other sources as possible is
needed and could be provided for by 
CP violation studies with $B$ decays and various rare decays like
$K^+\to\pi^+\nu\bar\nu$,
$B\to X_s\gamma$, $B\to X_s\nu\bar\nu$ or $B\to X_s\mu^+\mu^-$.
\end{itemize}

In our presentation we have largely focussed on such decays that
can be calculated reliably. In this spirit one may group the
various observables, roughly, into classes according to their
theoretical `cleanliness':
\begin{itemize}
\item
Class 1 (`gold plated'): $K_L\to\pi^0\nu\bar\nu$;
$K^+\to\pi^+\nu\bar\nu$, $B\to X_{s,d}\nu\bar\nu$
\item
Class 2 (very clean): $B(B_d\to l^+l^-)/B(B_s\to l^+l^-)$;
$\Delta M_{B_d}/\Delta M_{B_s}$
\item
Class 3 (moderate uncertainties and/or improvements possible):
$\varepsilon$, $B\to X_s\gamma$; $K_L\to\pi^0 e^+e^-$,
$B\to X_s l^+l^-$
\item
Class 4 (large hadronic uncertainties):
$\varepsilon'/\varepsilon$, $K_L\to\mu^+\mu^-$
\end{itemize}

The quantities that have already been measured, $\varepsilon$,
$B\to X_s\gamma$, $K_L\to\mu^+\mu^-$, or that are about to be
observed ($\varepsilon'/\varepsilon$, $K^+\to\pi^+\nu\bar\nu$)
are seen to cluster mainly in the lower part of this list.
Let us hope that future experimental developments will eventually
map out the full range of possibilities, including the unique
instances where unambiguous and clear theoretical predictions
can be made.

\section*{Acknowledgements}

I thank Andrzej Buras and Markus Lautenbacher for the very
enjoyable collaboration in ref. \cite{BBL}, on which much of the material
presented in this talk is based. I am grateful to the organizers
of the HQ '96 workshop in St. Goar for the invitation to this meeting.
Thanks are also due to Yuval Grossman for a critical reading of
the manuscript.

\end{document}